\documentclass{aa}  

\usepackage{natbib}
\usepackage{graphicx}
\usepackage{txfonts}
\usepackage{dsfont}
\usepackage{hyperref}
\usepackage{ulem}
\usepackage{enumitem}

\begin{document}

\title{The role of stochastic and smooth processes in regulating galaxy quenching}

   \author{Rain Kipper \inst{1}
          \and
          Antti Tamm\inst{1}
          \and
          Elmo Tempel\inst{1}
          \and
          Roberto de Propris\inst{2}
          \and
          Punyakoti Ganeshaiah Veena\inst{1,3}
          }

   \institute{Tartu Observatory, University of Tartu, Observatooriumi 1, 61602 T\~oravere, Estonia\\
      \email{rain.kipper@ut.ee}
   \and
   FINCA, University of Turku, Vesilinnantie 5, 20014, Turku, Finland
\and
Tata Institute of Fundamental Research, Homi Bhabha Road, Mumbai 400005, India
          } 
   \date{Received September 15, 1996; accepted March 16, 1997}

  \abstract
    {
    Galaxies can be classified as passive ellipticals or star-forming discs. Ellipticals dominate at the high end of the mass range, and therefore there must be a mechanism responsible for the quenching of star-forming galaxies. This could either be due to the secular processes linked to the mass {and star formation} of galaxies or to external processes linked to the surrounding environment. However, the contribution from these {smooth and stochastic} processes  to galaxy quenching has yet to be quantified.}
   { In this paper, we analytically model the processes that govern galaxy evolution and quantify their contribution. The key advantage of our method is that we do not assume the strength of the contribution from any of these processes beforehand, but instead aim to find their efficiencies. We have specifically studied the effects of mass quenching, gas stripping, and mergers on galaxy quenching.
    }
   {To achieve this, we first assumed a set of differential equations that describe the processes that shape galaxy evolution. We then modelled the parameters of these equations by maximising likelihood. These equations describe the evolution of galaxies individually, but the parameters of the equations are constrained by matching the extrapolated intermediate-redshift galaxies with the low-redshift galaxy population. In this study, we modelled the processes that change star formation and stellar mass in massive galaxies from the GAMA survey between $z\approx0.4$ and the present.} 
 {
  {We identified and quantified the contributions from mass quenching, gas stripping, and mergers to galaxy quenching. By modelling mass quenching, we found that quenching begins for galaxies above a mass of $\approx 10^{10.2} \; \rm{M_{\odot}}$, but is dependent on the gas accretion rate before quenching. The quenching timescale is on average $1.2$~Gyr and a closer look reveals support for the slow-then-rapid quenching scenario. The major merging rate of galaxies is about once per $10$~Gyr, while the rate of ram pressure stripping is significantly higher. In galaxies with decreasing star formation, we show that star formation is lost to fast quenching mechanisms such as  
  ram pressure stripping{ and is countered by mergers}, at a rate of about $41\%~{\rm Gyr^{-1}}$ and to mass quenching $49\%~{\rm Gyr^{-1}}$. Therefore, slow quenching mechanisms have a greater influence on galaxies in group or cluster environments than fast quenching mechanisms.
  }
  }
  {}

   \maketitle
\section{Introduction}
Galaxy formation and evolution are governed by a cocktail of physical processes, which are difficult to disentangle and quantify. One major obstacle is understanding and characterising star formation quenching, which depends on both galaxy mass and environment.

The bi-modality in the colour--magnitude diagram indicates that there are two distinct types of galaxy populations: blue star-forming spirals and red quenched ellitpicals (e.g. \citet{Baldry:2004, Schiminovich:2007}). Higher density regions in clusters contain higher fractions of E and S0 galaxies and far fewer star-forming spirals, which is known as the morphology-density relation \citep{Oemler:1974, Dressler1980}. The root cause can be twofold: {either the galaxies started off as they currently are in clusters and fields} or they have {evolved this way due to their interactions with the host environment.} This is {famously} referred {to} as {the} {nature versus nurture} of galaxies. {Unfortunately, it is almost impossible to ascertain the initial state of galaxies because this information is obscured: we cannot yet observe galaxies at very high redshifts, especially when they have still not yet formed stars
\footnote{{{
The appearance of a galaxy is attributable to the combination of its initial conditions at birth (nature) and its evolution history (nurture); for isolated galaxies, the effects of nurture are less evident than the effects of nature.
}
This approach as been pursued by e.g. \citet{AF:2015, AF:2016}}}.}
Therefore {the alternative is to investigate the {nurture} aspect and } study {environmental effects} as precisely as possible. The {main} aim of this paper is to {disentangle the environmental effects from each other and from secular processes and to quantify their contributions to galaxy quenching. Quenching can be due to galaxy mass, that is, a galaxy is so massive that further gas accretion and subsequent star formation are suppressed, or to several external processes such as ram pressure stripping that remove gas from a galaxy.} 

{However, to isolate and quantify the effects on galaxy quenching from environmental processes, we can rely on the time derivative of certain properties as they are {weakly} correlated with the environment compared to the correlation between galaxy properties and environment. }{For example, in the seminal work of \citet{Peng:2010} the authors studied mass and environmental quenching by estimating the fractions of passive and star forming galaxies as a function of either mass or environment. Studying these fractions gives the  current day estimates, but the important details are captured in the} time derivative of these fractions as they contain information about the evolution. This {implies} that studying the changes in galaxies over time provides opportunities to separate the effects of \textit{nature} {from those of} \textit{nurture}. 

{Galaxy properties evolve with time because of secular evolution and also because of external environmental processes} (e.g. \citet{jian2020redshift}). Both the secular effects and environmental effects can be described as a combination of {several} processes {(see a review of secular processes \citealt{Kormendy:2004})}. The processes that are considered secular are \textit{in situ} star formation \citep{santistevan2020growing}, gas accretion \citep{chen2020keck}, decay of star formation \citep{Maier_2019}, and over-consumption \citep{McGee_2014}. {A few examples of} environmental processes are accretion of \textit{ex situ} stars via minor mergers \citep{Di_Teodoro_2014, Deason_2016, suess2020color}, ram pressure stripping \citep{Gunn:1972}, galaxy merging \citep{Lotz:2011}, and even on whether they are main galaxies or satellites \citep{Peng_2012}. {This distinction} is not {very} strict, for example gas accretion rates can {also} depend on the environment.

In the present paper we present a galaxy evolution model, particularly concentrating on the  processes contributing to galaxy quenching, most of which {can be detected} in the {stellar} mass $M(t)$--SFR $\Psi(t)$ plane as a change in their position over time: 
\begin{enumerate}[label=\alph*)]
    \item \textit{In situ} star formation: production of zero-aged stars from the gas in the galaxy.
    \item  \textit{Ex situ} accretion of stars: increase of stellar mass that does not influence SFR. 
    \item {The rate of star formation decreases if the star formation rate  is high. This is because gas is used up to make stars, leaving no more gas available to make more stars.} 
    \item {Gas accretion: replenishes star formation and hence counterbalances the decay of SFR.} 
    \item {Galaxy mergers:} {depending on the type of merger, there can either be a sudden increase or decrease in the SFR.} 
    \item Ram pressure stripping: {decreases star formation} without affecting {the} stellar mass.
\end{enumerate} 

{Although} each of these processes can be described individually, their relative contributions can be disentangled only if modelled self-consistently and simultaneously. As a first approximation, all these {processes} can {be} described either as a first-order differential equation or as a discrete change of a parameter. 
Processes that {can be} approximated as first-order differential equations are {termed}  as {smooth} evolution {(processes a, b, c, d)} while discrete changes are referred to as {stochastic} evolution (processes e and f). Although we treat \textit{in situ} SFR as smooth evolution, it is expected to contain some randomness, for example, $0.17$\,dex in the case of a Milky Way-sized galaxy \citep{s2020stochastic}. {The mass quenching is {due to}  processes  (a), (c), and (d) given above, or the lack of these. { Gas stripping is due} to process (f), although tidal stripping can also contribute. }

A process not mentioned in the above list is the feedback from active galactic nuclei (AGNs). The precise influence of an AGN on a galaxy is not fully understood (e.g. \citet{Chen:2020}). The feedback to galaxy growth is expected to be low, as the supermassive black hole (SMBH) mass and the star forming disc mass do not correlate, but correlations exist with classical bulge \citep{Kormendy:2013}. There is a correspondence between AGN activity and star formation \citep{Aird:2019}. Here we assume that the influence of AGNs is  equivalent to delaying the star formation and its description is similar to stochasticy of star formation, hence smoothed over time. In the subsequent papers 
, we will further dissect the effects of AGNs. 

Processes that shape galaxy properties are modelled using three main techniques: analytic methods, semi-analytic methods, and numerical methods. Semi-analytic and numerical methods both rely on cosmological simulations and have successfully converged on a set of physical processes
that determine galaxy evolution \citep{Somerville2015}. However,
with the advent of large and deep sky surveys, it is now feasible to develop more accurate analytic models that reflect the  contribution of various processes that shape galaxy evolution in the Universe. In this work, we have established one such analytical model that also includes {the contribution to galaxy quenching from} stochastic events { in addition to}  smooth processes. 

The main goal of the study is to disentangle the various processes that contribute to galaxy quenching. We achieve this by tracing the growth of galaxies with simple equations that can be used to compare galaxies at different redshifts by mapping onto the same redshift\footnote{A similar method based on simulation data is called dynamical mode decomposition.}. We have successfully modelled and quantified the contributions
to galaxy quenching from the processes (a)-(f) listed above.

The paper is constructed as follows: in section 2 we describe  the analytic method that we have developed and implemented in this study, and in section 3 we describe the details of preparing the data, and validation of our modelling. Finally, in section 4 we present and discuss our results, followed by conclusions and scope for future work in section 5.

\section{Stochastic modelling of galaxy evolution}
\begin{figure*}
    \centering
    \includegraphics{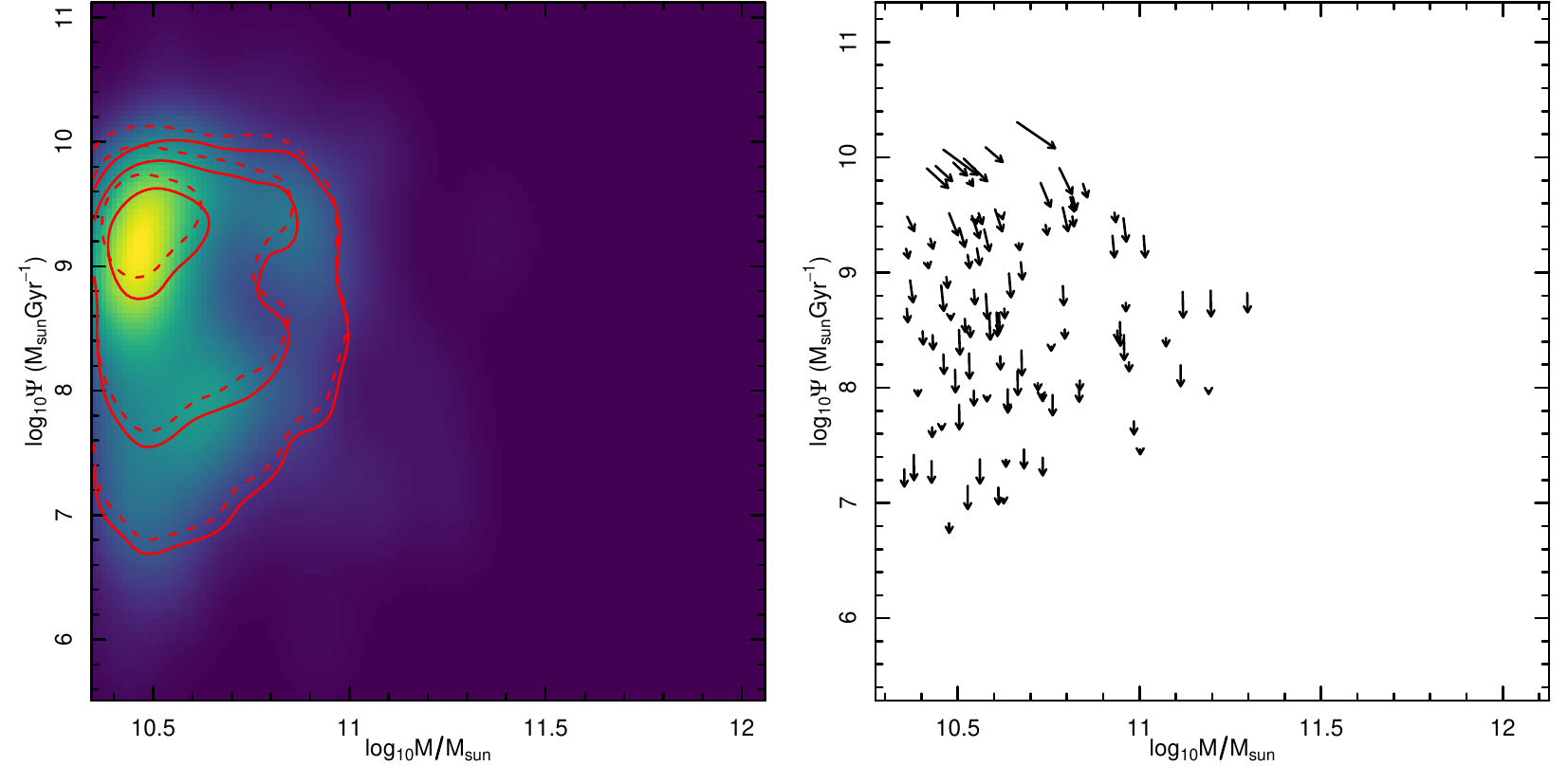}
    \caption{ Illustration of the method {in the stellar mass and SFR plane}. In the left panel, the dashed red lines show the distribution of galaxies at an intermediate redshift. The solid red lines show the distribution of galaxies at slightly lower redshift. {The colour in the background shows smoothed and continuous distribution of these points}. We infer that when constructing an evolution criteria for galaxies (shown as arrows on right panel) that matches the change of distributions on the left panel, we can infer the evolution of galaxies. }
    \label{fig:method_illustration}
\end{figure*}

We model the processes that shape galaxy evolution based on both stochastic events and continuous processes. This is accomplished by following the evolution of the galaxy population by inferring and remapping galaxies from higher to lower redshifts.
An illustration of this method is shown in Fig.~\ref{fig:method_illustration}. In principle, the idea of considering the observables of the whole galaxy population in order to study the impact of  only a {few} process on galaxy evolution was implemented by \citet{Drory_2008}. However, their focus is on the impact of galaxy mergers on the evolution of the galaxy mass function over cosmic time, while the goal of this article is to follow the growth of individual galaxies and then compare the entire population of evolved galaxies to those at another redshift. This is achieved by first writing down the equations describing the growth and then comparing the two distributions of galaxies with each other. In this section we describe the method. 

\subsection{General assumptions} \label{sec:math:non-st}

We distinguish two {distinct} families of galaxy evolution processes. {The first refers to the} gradual {growth such as}  gas accretion, {decrease in} star formation, gas enrichment, secular dynamical evolution, and so on. {In contrast, the second family refers to} stochastic processes like galaxy mergers and gas stripping which can change galaxy properties {instantaneously}. In order to model the evolution of a representative set of galaxies as a whole, both families of evolution processes have to be considered. 

As {discussed} in the introduction, the evolution of a typical galaxy can be described {by} a simple functional form, at least to a first approximation. {Out first assumption is that a galaxy property ${\bf x}$ evolves according to a first-order differential equation,}
\begin{equation}
    \frac{{\rm d}{\bf x}}{{\rm d}t} = f({\bf x},t, {\bf \theta}), \label{eq:general_evolution}
\end{equation}
where $f$ is the function {describing the evolution of the property with respect to}  time $t$, {and model parameter} ${\bf \theta}$. The function $f$ can be stochastic. {In this paper,} vector ${\bf x}$ consists of two fundamental galaxy properties, stellar mass ($M$) and star formation rate (SFR), both of which can be reasonably well inferred from observations. Nevertheless, in principle, this can be any other galaxy property as well, for example spin, metallicity, morphology, and others.

In the present study, our model of galaxy evolution includes crude estimates of the following processes: steady state star formation, star formation decay due to gas depletion, accretion of gas from the intergalactic medium, direct accretion of stars or small satellite galaxies, ram pressure stripping of gas, and major mergers. This list is by no means exhaustive, but encompasses all first-order effects. In the {subsequent} sections we describe {these} processes  and discuss how they affect {the specific} galaxy properties (${\bf x}=\{M,\Psi\}$): stellar mass ($M$) and star-formation rate ($\Psi$). In Sections \ref{sec:math:non-st} and \ref{sec:math:st} we describe the evolution of $M$ and $\Psi$. The non-stochastic processes are combined to a single set of differential equations by summing the contribution of each process. For example, increase in stellar mass of a galaxy is the sum of the amount of gas converted into stars and the steady accretion of stars by  merging with relatively much smaller galaxies. The non-stochastic part of the model is implemented by a sudden change of galaxy parameters happening at random times. Thus, a galaxy is assumed to follow a smooth evolutionary trajectory until some stochastic event (e.g. a merger with another galaxy) suddenly shifts its location in the parameter space, from where the smooth, non-stochastic evolution continues until the next stochastic event. We note that since we do not consider the chemical evolution and the evolution of the environment, the same age is attributed to all equal-mass galaxies.

\subsection{Non-stochastic processes} \label{sec:math:non-st}
\begin{figure}
    \centering
    \includegraphics[width=\columnwidth]{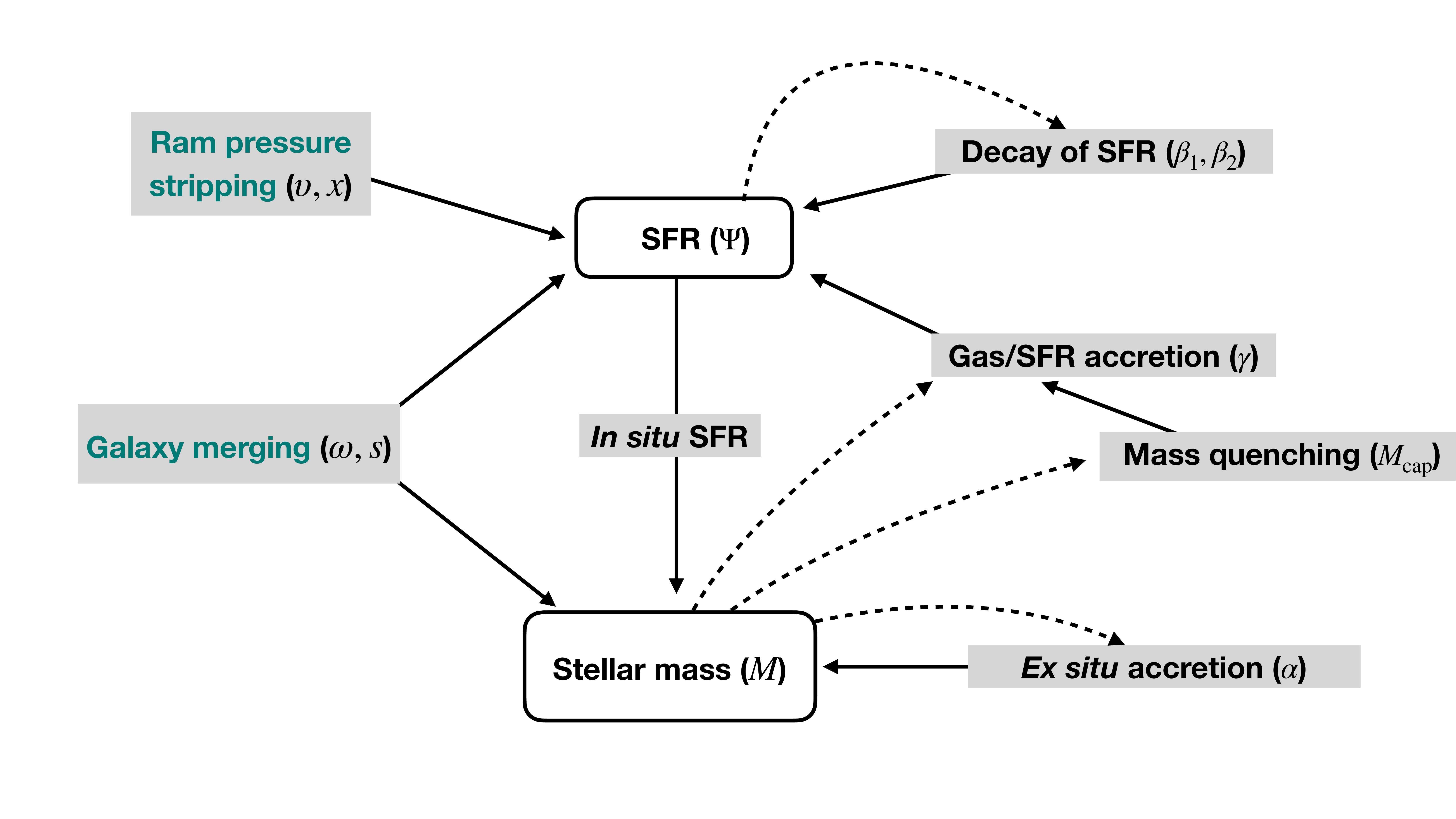}
    \caption{{Schema showing the relations between observables (in boxes) and processes (texts with grey background). The green text within the grey box indicates that the process is stochastic and black indicates it is smooth. The solid arrows show which of the observables are influenced by these  processes, and dashed arrows show that the observable galaxy parameter influences the process itself.} }
    \label{fig:schema}
\end{figure}

The differential equations describing the smooth evolution of galaxies are constructed by combining the effects of different processes on shaping galaxy properties:
\begin{eqnarray}
\frac{{\rm d}M}{{\rm d}t} &=& \Psi + \alpha M \label{eq:evol_M},\\
\frac{{\rm d}\Psi}{{\rm d}t} &=& -\beta_1 \Psi^{\beta_2} + \gamma M H(\log_{10}(M) - \log_{10}(M_{\rm cap})).\label{eq:evol_S}
\label{Eq=:smooth_evol}
\end{eqnarray}
The first part of Eq.~\eqref{eq:evol_M} describes the production of new stars from gas. The average brightness of newly formed stars is bright and the colour is blue, and so this process is directly observable as star formation rate. The second part of Eq.~\eqref{eq:evol_M} describes stars formed \textit{ex-situ}, and subsequently accreted to the galaxy. The extent of the ability of a galaxy to accrete stars is likely 
to depend on the gravitational potential, which is proportional to the mass of the galaxy. {These \textit{ex-situ} stars can be unbound, such as the sources of intracluster light or bound such as dwarf galaxies in minor mergers}. The strength of this process is reflected by the coefficient $\alpha$. The processes governing the evolution of SFR in Eq.\eqref{eq:evol_S} are divided into two parts. The first part describes the depletion of star formation due to the depletion of gas: the decrease of $\Psi$ is proportional to the amount of gas being depleted by star formation to some power (see Appendix \ref{sec:app_SF_derive} for a thorough justification of this form). The second part of Eq.~\eqref{eq:evol_S} describes the combination of two processes: gas accretion and mass quenching. 
The \textit{ex situ} accretion of stars is proportional to the mass of the galaxy, and this is represented by the accretion coefficient $\gamma$.  {Although Eqs.~\eqref{eq:evol_M} and \eqref{eq:evol_S} do not contain time explicitly, their time dependence is via their variables ($M$ and $\Psi$). } {A schema is shown in Fig.~\ref{fig:schema} to illustrate these processes. }
The Heaviside function describes how the ability of the galaxy to accrete gas is linked to the mass of the galaxy: whether the galaxy has reached its mass quenching limit ($M_{\rm cap}$) or not. 

\subsection{Stochastic evolution} \label{sec:math:st}

In group and cluster environments, we can envisage two major processes which substantially change galaxy properties over a relatively short time interval: (a) gas `stripping' (ram-pressure or tidal) and (b) a major merger with another galaxy with stellar mass that is equal to between one-quarter and one times that of the primary galaxy. Mergers with lower-mass galaxies are considered as the `dry' accretion, as specified above. In our modelling, the effect of stripping is the multiplication of star formation rate by a factor $0 < x \leq 1$, such that 
\begin{equation}
    \Psi = x\Psi_0  .\\
\end{equation}
Here, $\Psi_0$ denotes  the SFR before the stripping event and $\Psi$ denotes  after. In the long term, gas is  stripped, and therefore 
star formation decreases heavily \citep{Kenney:2014}. In some cases such as the jellyfish galaxies, a short-term SFR elevation is observed, because during the stripping event, gas is compressed leading to star formation
\citep{Vulcani_2018, Vulcani:2020}. 

Our aim is to model the more long-term effects, and so we set $x=0.8$, meaning each ram pressure event strips $20\%$ of the gas and  accordingly reduces star formation. Fixing this value is a matter of defining a ram pressure stripping event and is somewhat arbitrary, as it strongly depends on the geometry of the stripping event such as: an edge-on versus face-on movement with respect to gas, the relative velocity of the intergalactic medium and galaxy, and relative gas densities. Differences between 
these geometrical effects will be compensated 
by possibility there are many events
. Fixing the value of $x$  very high would increase the number of these events, while a lower value would underestimate this. We tested {our model} by fixing $x=0.6$ and $x=0.8$ { and} in both cases the cumulative effect remained the same (less severe but more frequent ram pressure stripping would produce the same results as more severe but less frequent ram
pressure stripping).  ` \citet{ramatsoku2020gasp} detected a `jellyfish' galaxy with a long tail, allowing us to make a comparison. The tail was formed when the jellyfish galaxy had a velocity deviation of $\sim 3.5\sigma$ , which is quite an extreme velocity for a galaxy in a group. This galaxy had $60\%$ less gas than expected based on its stellar mass. Considering that $60\%$ gas removal is an extreme event, we define ram pressure stripping as an event removing $20\%$ of the gas and accordingly reduces star formation. 

{The increase of stellar mass and SFR during merger events is characterised by the following equations}
\begin{eqnarray}
    M &\rightarrow& M + M_{\rm merger}\\
    \Psi &\rightarrow& (\Psi + \Psi_{\rm merger})\times s.
    \label{Eq=:stoch_evol}
\end{eqnarray}
{We denote the SFR and mass of the merging galaxy with the index `merger'. In cases where the merging galaxy has a higher mass compared to the galaxy currently being evolved, then the galaxy is removed from the sample. }
In eq.(6), $s$ is a starburst coefficient, enabling us to take into account the rapid enhancement of star formation rate during mergers that is evident from observations \citep{Cortijo:2017, DiazGarcia:2020}. 

The stripping and merging events are timed randomly over the modelled time interval. The frequency of the events is assumed to be a Poisson process and is determined from the modelling. We denote the rate of stripping events as $\upsilon$ or $\upsilon(x=0.8)$, and the rate of mergers as $\omega$. 

Each merging and stripping event can have a huge impact on the eventual properties of a given galaxy. In the case of a limited galaxy sample, the uncertainties of such stochastic events will render the properties of the overall galaxy population uncertain. We can overcome such `shot noise' with a statistical trick, applying the stochastic evolving algorithms 
many times to each galaxy over a given time interval and calculating the net distribution of $M$ and $\Psi$. The possibility that the galaxy ceases to exist is also represented in this net distribution. This repetition reduces the noise, which is needed for likelihood stability in modelling. {A general caveat here is that realisation and expectation are
treated as the same thing. A mock data analysis shows that this is not
relevant for the present case.} For a large dataset, this procedure is not necessary. 

\subsection{Statistical inference} \label{sec:statistical_implement}
Provided with a sufficiently large observational dataset of galaxies with properties $M$ and $\Psi$ over a sufficiently large range of redshifts, we can let each observed galaxy evolve in the $M$ and $\Psi$ space according to the prescriptions mentioned above. From the redshift, we can infer the time $t$ at which we are seeing a given galaxy. This defines the starting point of galaxy evolutionary trajectory. The exact point of zero time is inconsequential as Eqs.~\eqref{eq:evol_M} and \eqref{eq:evol_S} do not use time explicitly. Using the formed ensemble of trajectories of all the galaxies, we can calculate the probability density function (PDF) 
of the properties $M$ and $\Psi$ at any moment $t_a$. By comparing these PDFs to the parameters of the galaxies actually seen at the time $t_a$ we can evaluate the likelihood of the parameters chosen for {Eqs.}~\eqref{Eq=:smooth_evol}--\eqref{Eq=:stoch_evol}.

For practical considerations, we need to smooth the PDF with a kernel $K$. In the following, we denote each observed galaxy as $y_i$ and its parameters at the time of the observation as ${\bf x}_{t}(y_i)$. 
We consider each individual galaxy $y_i$ to belong to the set of all galaxies, $y_i\in Y$. For each galaxy, its parameters $M$ and $\Psi$ are denoted ${\bf x}_{t}(y_i)$, where $t$ denotes the time at which the galaxy is currently being seen.

The core of statistical inference lies in finding and maximising the likelihood. Likelihood calculation requires data, and the probability density function. For the present case, we find the probability density function by combining extrapolated data at the time of likelihood evaluation. We denote the PDF $p({\bf x}|t){\rm d}{\rm x}$. As galaxies and therefore the PDF evolves, we must always specify the time at which the PDF is evaluated -- $t$. In the present approach, we find the PDF by extrapolating earlier galaxies to each time where we evaluate likelihood. We denote the extrapolation of all galaxy $y_j$ properties from $t_j$ to time $t_i$ as ${\bf x}_{t=t_j}^{t=t_i}(y_j)$. This extrapolation is done by solving (numerically) Eq.~\eqref{eq:general_evolution} with initial conditions ${\bf x}_{t_j}(y_j)$. The conversion from extrapolated points to smooth PDF is done using kernel $K$:
\begin{equation}
    p({\bf x}|t_i) = Z^{-1}\sum\limits_{j\ne i} K\left({\bf x}, {\bf x}_{t=t_j}^{t=t_i}(y_j)\right)\label{eq:kernel}.
\end{equation}
We denote the normalisation constant of the PDF as $Z$. In the present application, the kernel is implemented using a randomly located grid with 
\begin{equation}
    K({\rm x, x'}) = \mathds{1}(\{{\rm x,x'}\}\in g), \label{eq:grid_kernel}
\end{equation}
where $g$ denotes grid cells. 
In Eq.~\eqref{eq:kernel} we find the PDF for the point at time $t_i$. The summation is done over other galaxies, except the one that is used to evaluate the likelihood. This is needed to keep the data and model independent. During the evaluation of the likelihood at $t_i$ the point $y_i$ is data, and all other points and their extrapolations are considered as model. The above-mentioned condition holds if the data are independent and identically distributed. If the evolution of the galaxies can be reversed (i.e. calculating the appearance of galaxies when we de-evolve them), then summing is done over galaxies where $y_i\ne y_j$. If the galaxy evolution cannot be reversed (e.g. when there is merging/stochastic processes), then summation is over $j$ where  $t_j<t_i$. 

The likelihood for the modelling is evaluated over all data points:
\begin{equation}
    \log\mathcal{L} = \sum\limits_{i=1}^{N} \log p[{\bf x}_{t=t_i}(y_i)|t_i ]. 
\end{equation}
The maximum of the likelihood gives the values to model parameters ${\bf \theta}$.


\section{Data and implementation}
\subsection{Data}
In order to apply the above method we need an observational dataset providing the masses and star formation rates for a large sample of galaxies over a cosmologically significant redshift range. In addition, we need to know whether a given galaxy is in a group environment or not.

The Data Release 3 of the Galaxy and Mass Assembly Survey \citep[GAMA;][]{GAMA:DR3} provides a suitable dataset for our study. It is complete down to the $r$-band magnitude $19.0^m$ or $19.4^m$ (depending on the used fields G09, G12, G15), and we can use it to reach $z\simeq0.5$, corresponding to an evolution time-span of up to $\sim5\,{\rm Gyr}$. The stellar masses and SFRs are derived by \citet{Driver:2018} with the MAGPHYS algorithm \citep{daCunha:2008} and also using the Herschel Space Telescope far-infrared observations and UKIRT near-infrared observations, thereby forming a strong basis for realistic SFR and stellar mass estimates. 

The group environment was probed by \citet{Robotham:2011} with the popular and much-tested Friends-of-Friends method \citep{grvordlusartikkel}, finding that the sample contains $15\,545$ isolated and $14\,882$ group galaxies (including galaxies in pairs). However, we note that the authors promote caution, as there is a noticeable deficit in the number of rich systems compared to their mock catalogue analysis.

We use the redshift to ascribe the variable $t_{\rm evol}$ to each galaxy, indicating the time since the pre-defined (but generally arbitrary) starting point, assuming the cosmological parameters $H_0 = 71\,{\rm km\,s^{-1}\,Mpc^{-1}}$, $\Omega_M = 0.27$, $\Omega_\Lambda = 0.73$. 

\subsection{Implementation}

As galaxy masses generally increase over time, an ideal dataset would have a lower mass limit decreasing with increasing redshift, just the opposite of the effect of the Malmquist bias. For the modelling, mass completeness of the unit evolution cells is crucial; incompleteness would lead to spurious evolutionary paths. The GAMA sample is originally flux-limited, and therefore in order to be able to properly model the mass evolution we had to construct narrower redshift bins, hereafter referred to as `unit evolution cells' containing locally complete subsamples (in stellar mass), and study the evolution between consecutive cells. The likelihood was calculated for each unit evolution cell separately and overall likelihood was found by multiplying them.  For the actual cell construction, we had to find a compromise: narrower cells would enable us to make use of a larger fraction of galaxies and a broader mass range, but the time-span would be smaller, making it difficult to detect any evolution. 

\begin{figure}
    \centering
    \includegraphics{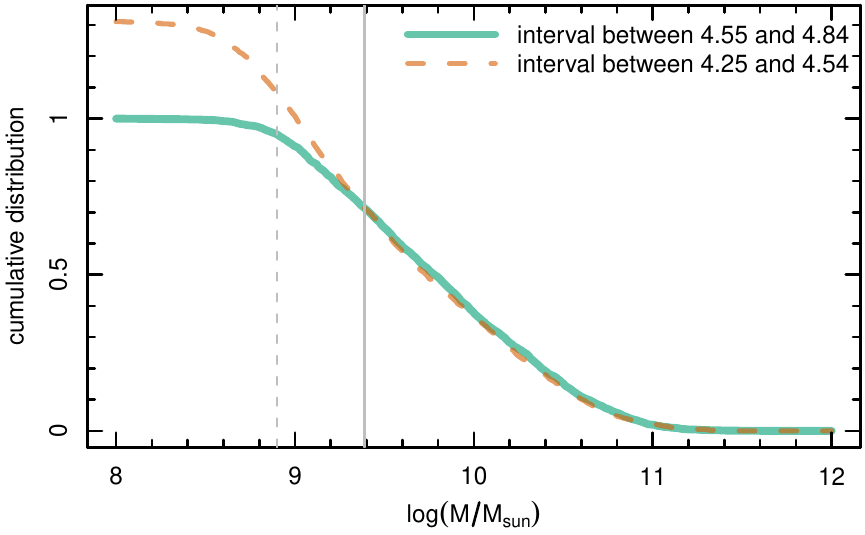}
    \caption{Modelling of the completeness of the dataset. The green and red lines show the cumulative mass distributions at two subsequent time intervals normalised to match at the high-mass end. The vertical grey line shows the mass above which the sample is complete and the dashed grey line shows the same for the previous time interval. 
    {These time intervals do not correspond to any specific time intervals in Fig.~\ref{fig:cope_with_fluxlim}.} }
    \label{fig:mass_completenss_illustration}
\end{figure}

\begin{figure*}
    \centering
    \includegraphics{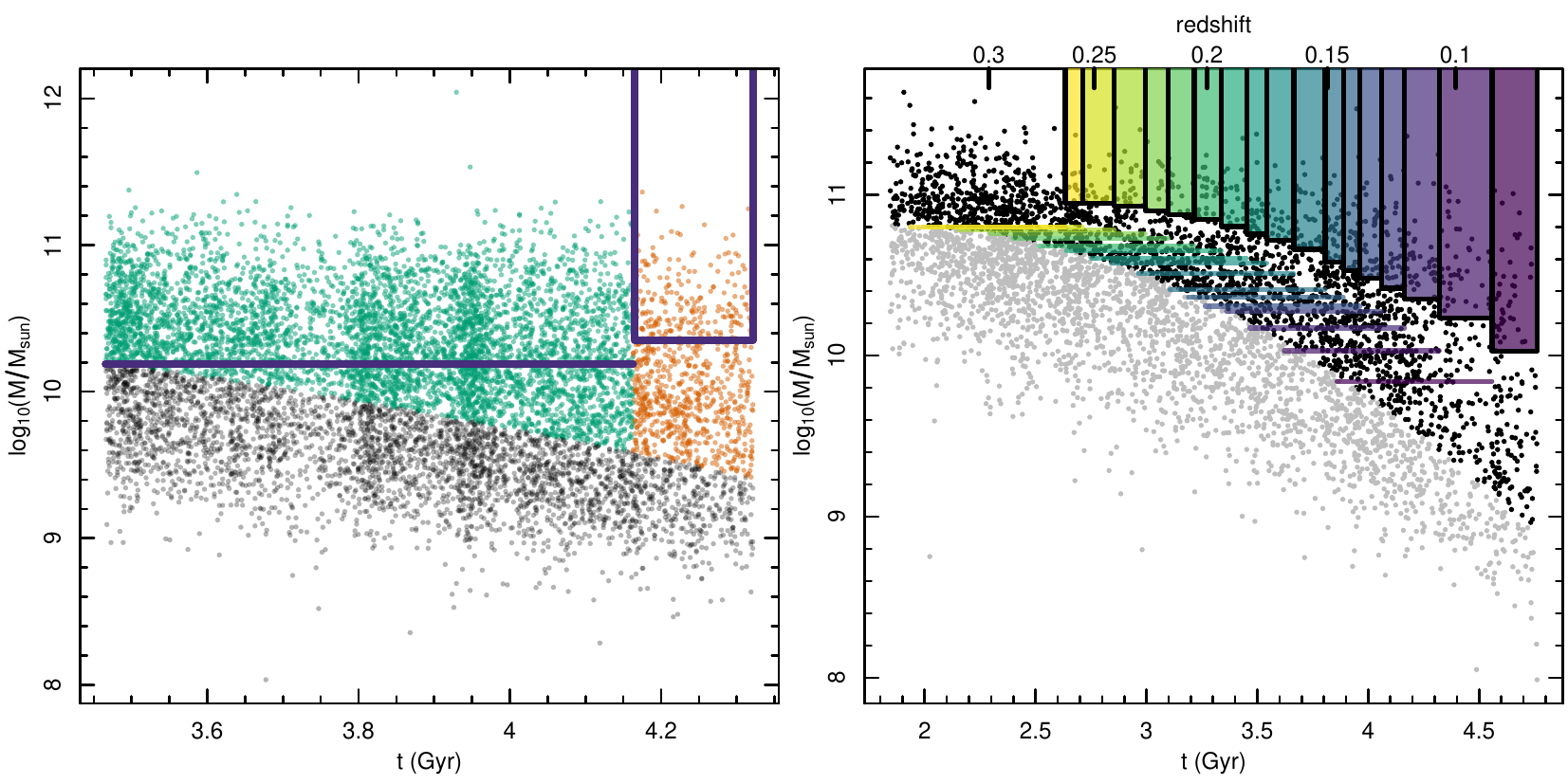}
    \caption{Construction of a locally volume-limited galaxy sample. In the {left {hand} panel}, an example of a small volume-limited sample is shown, which is used as {the unit sample set for modelling galaxy evolution. }
    The horizontal line indicates the redshift range of galaxies that are used to generate the PDF (see Sect.~\ref{sec:math:non-st}); the vertical positioning of the line shows the mass-completeness limit. The  {purple} box defines the region used for evaluating the PDF. The vertical gap $\Delta M_{\rm complete}$ between them shows the maximum evolution {that can be modelled } with this configuration {(see the main text for more explanations)}. The \textit{right {hand} panel} shows all the modelling elements denoted with different colours. In both panels, the grey points represent a (subsample) of the GAMA galaxies in that redshift range.
    }
    \label{fig:cope_with_fluxlim}
\end{figure*}

The unit evolution cell construction requires mass completeness. {We estimated the completeness} by comparing mass functions at different time intervals. We assume that flux limitedness has a stronger influence on the low end of the mass function than galaxy evolution has on the high end. By matching the high end of the mass function of neighbouring time intervals, we selected a low mass limit where these started to diverge. This diverging point is the mass limit within which the current flux-limited dataset is adequate
. An illustration of this divergence is shown in Fig.~\ref{fig:mass_completenss_illustration}. 
For the nearby universe starting time interval we obtained a value similar to \citet{GAMA:DR3}, the precise value being $\log_{10}M_{\rm complete}/M_\odot = 8.9$ at redshift $0.04$\footnote{\citet{GAMA:DR3} adopted a mass completeness of $\log_{10} M_{\rm complete}/M_\odot=9.0$ up to redshift $0.06$, which is 
{slightly higher than our redshift cut.}
}. We estimated the completeness of each subsequent time interval iteratively. The overall smooth completeness was found by fitting a second-order 
polynomial through these centres of intervals. This polynomial had the values $\log_{10}M_{\rm complete} = 10.07 + 0.81t - 0.22t^2$ and is seen in Fig.~\ref{fig:cope_with_fluxlim} as a separation between grey and black points.

Having determined the completeness limit,  the dataset must be divided into unit evolution cells. On one hand, a small extent in redshift or time of the unit evolution cells could ensure a uniform completeness across the whole redshift range, and fewer galaxies would have to be expelled from the sample, improving the statistics. On the other hand, a short time-span $\Delta t$ of each cell would make it difficult to render the evolutionary effects in our modelling (i.e. $\dot M\Delta t$ becomes too small compared to $M$), making the results uncertain. After some testing, we chose $\Delta t \sim 0.6~{\rm Gyr}$ for unit evolution cells and required that each cell contain at least $\approx 200$ galaxies to enable the evaluation of the likelihood in that cell. In the left panel of Fig.~\ref{fig:cope_with_fluxlim} we show an example of the unit evolution cell. The {green} points show the galaxies that are extrapolated in time direction. The { purple} horizontal line shows the adopted completeness limit for that cell. The {orange} points show galaxies that are in the region for likelihood evaluation of that cell. The { purple} top-open box shows the galaxies that are actually used. We introduced a gap between the completeness limit of that cell and the minimum mass of the galaxies used for likelihood evaluation in order to ensure that possible evolution would not require galaxies below the completeness limit (or the evolution determination would not be influenced by the Malmquist bias). We chose the gap size to be such that the evolution of the galaxy should not exceed $50\%$ per Gyr. To save computational time, we assumed the same PDF for galaxies at a similar evolutionary stage (see green points on the left panel of Fig.~\ref{fig:cope_with_fluxlim}). The right hand panel in Fig.~\ref{fig:cope_with_fluxlim} shows all the used unit evolution cells {with each horizontal line in the right hand panel corresponding to the green regions of galaxies in the left panel that are extrapolated. The vertical boxes in the right and left hand panels have one-to-one correspondence for each evolution cell}.

The crucial point of inference is the  adjustment for the similarity of higher and lower redshift galaxies. The similarity is measured with likelihood, which requires selecting kernel $K$ in Eq.~\eqref{eq:kernel}.  For this, we used the grid in Eq.~\eqref{eq:grid_kernel}: we find that filling of grid-cells of extrapolated and observed galaxies is proportional. For the implementation, the distances of galaxies and grid centres on the mass--SFR plane are calculated as $d = \sqrt{ (\log_{10}M - \log_{10}M_{k})^2 + b(\log_{10}\Psi - \log_{10}\Psi_{k})^2}$, where $k$ indexes the centres of Voronoi bins, and the weight factor $b$ is used for tuning the balance between mass and SFR. We use $b = 8$ for the isolated galaxy sample and $b=7$ for the group sample. Selecting larger values for the $b$ value causes the grid to be elongated in the mass direction. This approach was chosen intentionally so as to cause higher sensitivity in order to model the SFR evolution: we aimed to study quenching and  a finer grid in the quenching direction increases its sensitivity.  
A similar approach to likelihood evaluation but without reducing the grid to a vector was taken by \citet{Kipper_bartorque}. 

An advantage of binning compared to smoothing is that binning does not artificially broaden the PDF as convolution would. A drawback is that each time data are binned, some information is lost. In order to regain the information, we calculate the likelihood $L$ with different Voronoi cells and find the expectation of likelihood (the inference works for a single likelihood, and since the optimum parameter values are shared, the expectation of likelihoods is also bound to give the same  optimum). We tested this on a mock catalogue which gave viable results; see Sect.~\ref{sec:mock} for more information.  For the implementation, we used $35$ Voronoi cells and $500$ evaluations for smoothing the PDF and finding the expectations of the likelihood. These numbers insured smooth and stable likelihood evaluations. The process of maximisation was done with the Multinest algorithm  \citet{MN1, MN2, MN3}. We used $500$ live points and wide and uniform priors for inference. 

\subsection{Validation of the method on mock data}\label{sec:mock}
Before application of a new method on real data, a test of reliability can be beneficial. In order to test the reliability, we generated a mock dataset by evolving an individual seed galaxy from a random initial position according to the analytical formulae in Eqs.~\eqref{eq:evol_M}-\eqref{Eq=:stoch_evol}. The parameters describing the evolution were chosen randomly, but roughly within their expected range. We picked the resultant parameter sets $\{M,\Psi\}$ at random epochs to mimic a galaxy being observed at a non-determined redshift. 
\begin{figure*}
    \centering
    \includegraphics{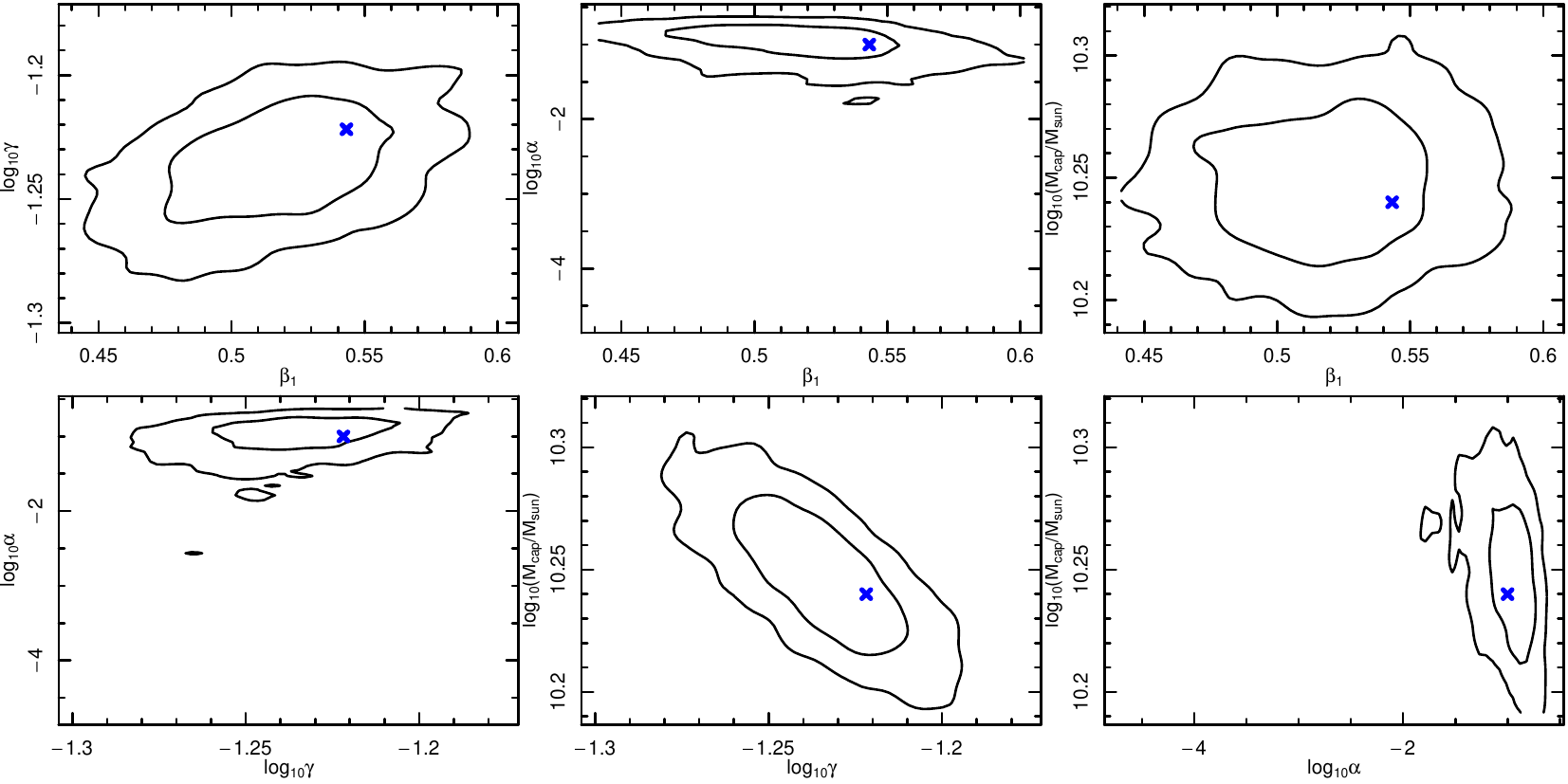}
    \caption{{Posterior distributions of mock data modelling for method validation. The contours show $1\sigma$ and $2\sigma$ confidence levels and the blue cross shows the true value with which the mock was generated. {Based on this we infer that } {our} modelling captures the true values well. }}
    \label{fig:mock_posterior}
\end{figure*}

The resultant mock catalogue was then used as input for the modelling algorithm in order to check the consistency of the method. The results are shown in Table~\ref{tab:mock_res_non_stochastic} {and Fig.~\ref{fig:mock_posterior}}. We can see that all the input parameters were recovered well within the uncertainties. 
\begin{table*}
    \centering
    \caption{Results of non-stochastic modelling using the mock data. The table shows variables used for mock generation, their true value, and the average value we acquired from modelling with their uncertainty. The shown tension is the difference between the fitted and true value over uncertainty. The Pcumul is the value of the cumulative distribution of the posterior at the true value; in case of perfect modelling, these values should be uniformly distributed between $0$ and $1$. The priors show search range in parameter fitting. We conclude that resulting modelling covers values accurately and without noticeable bias. }
    \label{tab:mock_res_non_stochastic}
    \begin{tabular}{l|lllllll}
    \hline
    \hline
    Variable & True value & Fitted mean & Fitted st.dev & Tension & Pcumul & Prior low & Prior high \\
    \hline
$\beta_1$ & 0.543 & 0.517 & 0.028 & -0.941 & 0.836 & 0 & 10\\
$\log_{10}\gamma$ & -1.222 & -1.236 & 0.017 & -0.829 & 0.801 & -5 & 2\\
$\log_{10}\alpha$ & -1 & -1.024 & 0.394 & -0.061 & 0.362 & -5 & 2\\
$\log_{10}(M_{\rm cap}/M_\odot)$ & 10.24 & 10.248 & 0.021 & 0.399 & 0.357 & 8 & 12\\
\hline
    \end{tabular}
\end{table*}

For testing the stochastic part, we fixed the non-stochastic part of the modelling (as we did in application), and reran the next mock catalogue. For simplicity, only the stripping was included. The resulting ram pressure stripping rate was recovered within $1.2$ standard deviations. {The fitted value for the stripping rate was $\log_{10}(\upsilon) = -0.32^{+0.06}_{-0.07}$ while the true value was exactly $-0.4$}. 

We conclude that a test on a mock catalogue showed that the modelling approach is able to recover the underlying evolution parameters. The test on the mock relies on the assumption that the functional form of the evolution is able to describe the true underlying evolution of galaxies.

\section{Results and Discussion}

In this section, we present the results obtained by the modelling described in Sect. 2. It is well established that galaxy evolution strongly depends on the environment. In order to model galaxy evolution more accurately and to include processes specific to the environment, we simply split the sample into isolated and non-isolated galaxies. The modelling results for these two cases are presented in separate sections. 

\subsection{Evolution in isolation}
Most of the galaxies in our sample appear to be on the threshold of quenching, and therefore we are able to impose the strongest constraints on the quenching parameters, $\beta_1$ and  $\beta_2$.  The timescale of quenching is described as 1/$\beta_1$ and its effectiveness by {unitless} $\beta_2$ compared to an exponential decay, which is a natural outcome of a closed-box model. 
\begin{figure}
    \centering
    \includegraphics{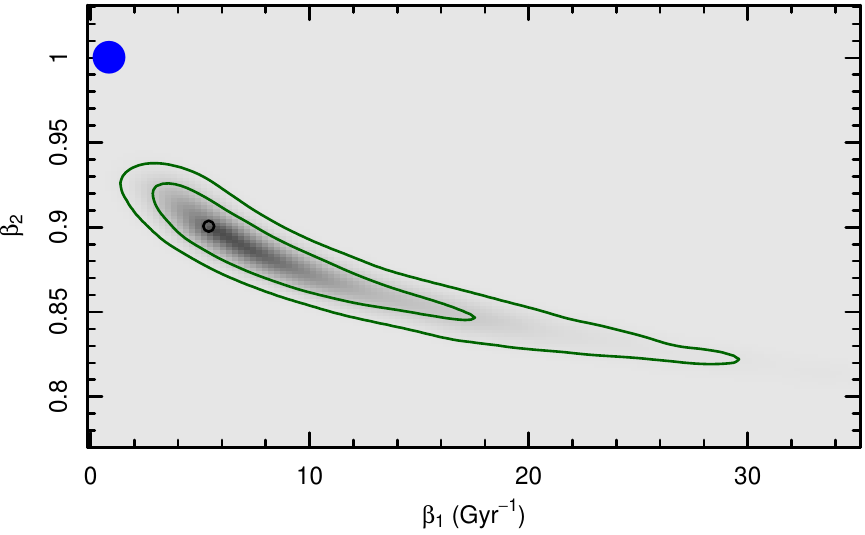}
    \caption{Decay of star formation described by the equation $\dot\Psi\propto -\beta_1\Psi^{\beta_2}$ (see Eq.~\eqref{eq:evol_S}) and its posterior distribution. The parameter $\beta_1$ depicts the speed at which galaxies quench; a higher value indicates faster quenching and $\beta_2$ describes the quenching behaviour and is related to the Kennicutt-Schmidt relation {(see appendix~\ref{sec:app_SF_derive})}. The green lines cover 65\% and 95\% of the total distribution. The large blue dot shows the modelling results for the case where  $\beta_2=1$. 
    The secular evolution values need to be fixed for studying environmental effects. For the fixing we used the values depicted with the black dot.  }
    \label{fig:decay_plane}
\end{figure}
For a straightforward comparison with the literature, we used a fixed value of $\beta_2=1$, obtaining an exponential timescale for quenching of $\tau \equiv 1/\beta_1 = 1.5\pm0.1~{\rm Gyr}$, which is comparable to the \citet{Moutard:2016} estimate of $\tau = [0.5, 2]~{\rm Gyr}$ and that of \citet{Haines:2015} of $\tau = 1.73\pm0.25\,{\rm Gyr}$ obtained from star formation analysis of galaxies falling into clusters. While keeping the $\beta_2$ free, we acquire its value less than one. The full posterior of the modelling is shown in Fig.~\ref{fig:decay_plane}. This means that the change in SFR is proportional to $\Psi^{0.9}$, not straight $\Psi$ as in the case of a closed-box model. For larger $\Psi$ values, the exponent causes the decay of SFR to be slower compared to a closed-box model. Once the $\Psi$ reaches small values then the reduction of $\Psi$ increases compared to the closed-box model. A similar behaviour of having a lower  rate of decay for higher SFR values was reported by \citet{Maier_2019} and exhibits a slow-then-rapid quenching. 

\begin{figure}
    \centering
    \includegraphics{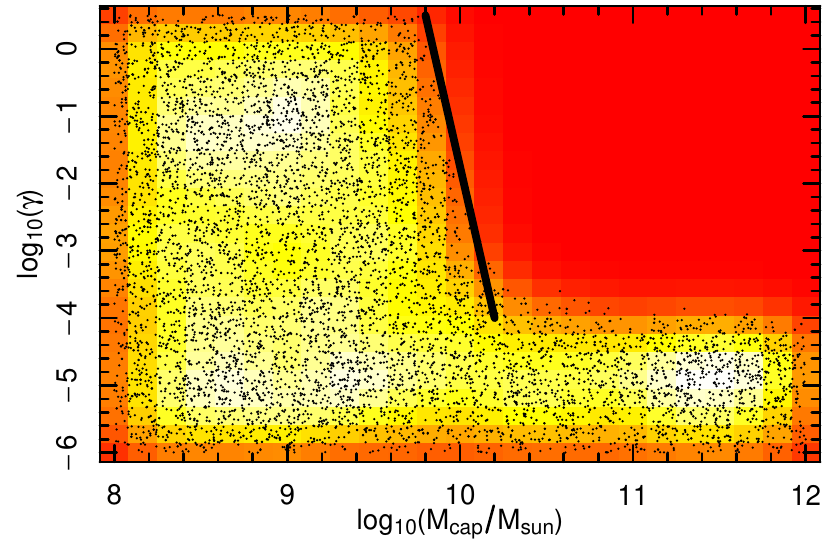}
    \caption{{The mass quenching determination dependence on the gas accretion. }
    The x-axis represents the mass of a galaxy at which it starts quenching, plotted against $\gamma$, which depicts the rate of gas accretion by the galaxy. Each dot represents a possible solution as posterior samples from the Multinest fitting. The colour background shows smoothed distribution of the posterior points. The solid black line given by ($\log_{10}\gamma = 110.75 - 11.25\log_{10}(M_{\rm cap}/M_\odot) $), shows that a galaxy that accretes gas {in high quantities} quenches  at a lower mass compared to a galaxy that does not accrete a lot of gas.}
    \label{fig:result_mass_cap}
\end{figure}

The model parameter $M_{\rm cap}$ corresponds to the galaxy mass at which quenching starts and is therefore an important indicator. Figure \ref{fig:result_mass_cap} suggests that it has a strong degeneracy with the parameter  $\gamma$, describing the efficiency of gas accretion. 
Each dot in \autoref{fig:result_mass_cap} shows a possible and an almost equally probable solution. According to the lower part of the figure, no quenching is required at all for low gas accretion values. We consider this result an artefact of the limited redshift and mass range of the galaxy sample. Another apparent conclusion from the figure is that all low quenching values are allowed. Again, this is a product of the lack of low-mass galaxies in the sample. Despite these limitations, we obtain a constraint on the mass at which a galaxy quenches for significant gas accretion. In such cases, quenching must start before a galaxy reaches a mass of $10^{10.2}~{\rm M_\odot}$, indicated with the black solid line in Fig.~\ref{fig:result_mass_cap}. This result is enhanced by the slow-then-rapid nature of the quenching as initially the decay of SFR is slow. We would like to point out that our result is quite close to the completeness limit (see Fig.~\ref{fig:cope_with_fluxlim}). 

\citet{Moutard:2016} found the mass cap to be $\log_{10}(M/M_\odot) = 10.64\pm0.01$. These latter authors determine the mass quenching limit by finding the turning point of the mass distribution function, which is somewhat different from our  approach, where the value corresponds to the start of quenching, and therefore a value larger than the one reported by \citet{Moutard:2016} is expected. {A similar value was found by \citet{Argudo_Fernandez:2018} based on AGN studies. \citet{Peng:2010} demonstrated different causes of quenching either by mass or environment. These latter authors showed that mass quenching starts from quite low masses. } 
In the EAGLE simulation,  \citet{Cochrane:2018} suggested that mass quenching is operative for galaxies with stellar mass higher than $\sim10^{10}\,{\rm M_\odot}$.

We also account for the possibility that not all stars present in a galaxy have been formed in situ, but could have been acquired from the surroundings. The model parameter $\alpha$ in Eq.~\eqref{eq:evol_M} controls the rate of such `dry' accretion. For isolated galaxies, the cumulative distribution of $\alpha$ values over all Multinest samplings is presented in Fig.~\ref{fig:dry_accretion}. Although a slight bimodality is present in the distribution, about $90\%$ of the samplings favour an accretion rate in the range of $0.18~{\rm Gyr^{-1}}$. This value suggests that in order to double the galaxy mass in isolation via minor mergers, it would take at least $\ln{2}/\alpha\approx3.8\,{\rm Gyr}$. Based on colour gradients, \citet{suess2020color} suggest that quenched galaxies do evolve, mainly via minor mergers.

\subsection{Evolution in groups}
\begin{figure}
    \centering
    \includegraphics{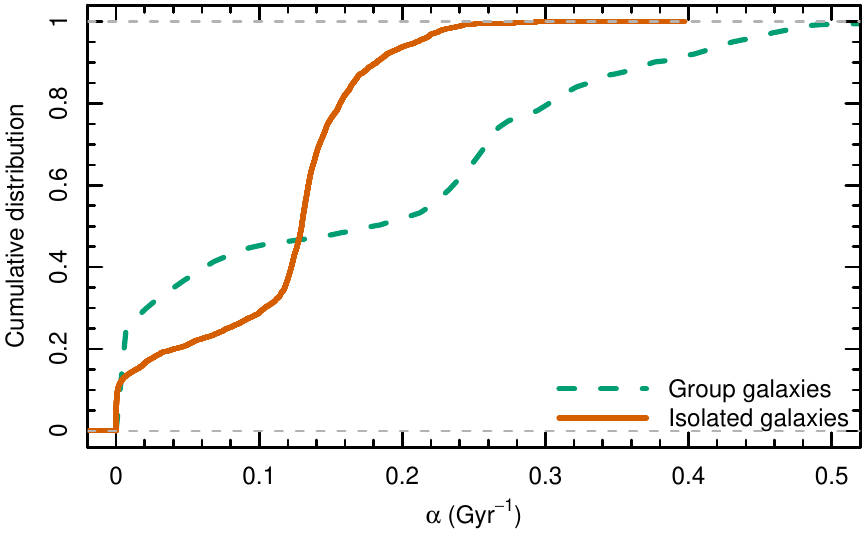}
    \caption{Cumulative posterior distribution of the $\alpha$ parameter, which describes the accretion of stars into a galaxies. The stars being accreted can be from the intergalactic medium, via minor mergers, or from globular clusters. The { dashed green} line shows the fitting results for group galaxies, while { the orange} line shows isolated galaxies. On average, the group galaxies accrete more stars compared to isolated ones.}
    \label{fig:dry_accretion}
\end{figure}
In denser environments, 
a few differences in galaxy behaviour are expected compared to the isolated galaxies. The parameters for which we anticipate differences (only $\alpha$) were fit once again  by incorporating  group-specific parameters. 

As expected, the `dry' accretion parameter performs somewhat differently in a group environment than for isolated galaxies, which is shown as the blue line in \autoref{fig:dry_accretion}. In general, dry accretion is elevated compared to isolated galaxies, but remains below $0.37~{\rm Gyr^{-1}}$ in $90\%$ of the solutions, meaning that in 1 Gyr, galaxy mass increases by less than $37\%$ via direct accretion of stars, for example, via minor mergers. In the case of minor mergers, we do not consider the increase of gas mass from minor mergers, as it is estimated to be a maximum of $0.28\,{\rm M_\odot\,yr^{-1}}$, which is much lesser than what is expected for stable star formation \citep{Di_Teodoro_2014}.

In denser environments, galaxies may also experience ram pressure stripping. Our modelling gave the value $\upsilon = 2.4 \pm 0.6~{\rm Gyr^{-1}}$ for the stripping rate, assuming that each event removes $20\%$ of star formation of the galaxy. For a consistency check, we tested the situation when each stripping event would remove 40\% of the SFR. As a result, the occurrence of stripping events was reduced by 50\%, as expected. 

On average, ram pressure stripping removes $1-0.8^{2.4}=41\%$ of star formation each gigayear for massive galaxies. This value is very similar to the mass quenching in isolated massive galaxies, which removes $1-\exp{(-1/1.5)} = 49\%$ of star formation in 1 Gyr. Assuming that the mass quenching mechanism is roughly similar in all environments, we see that \textit{in situ} and external quenching have a comparable importance in dense environments. {A simple application of these numbers is to estimate the length of time ($t_{\rm GV}$) that a galaxy would stay as a green valley galaxy which is in between the star forming region and quenched galaxies. When we include both environmental effects and mass quenching, and also consider the span of the green valley in SFR to be $0.1$~dex, we estimate the scale as
\begin{equation}
    x^{\upsilon t_{\rm GV}}\exp(-\beta_1t_{\rm GV}) = 0.1.
\end{equation}
Numerically the $t_{\rm GV}$ is $1.9$~Gyr. This is in agreement with \citet{Schawinski:2014} who have shown that disc galaxies quench over 1 Gyr.}

\begin{figure}
    \centering
    \includegraphics{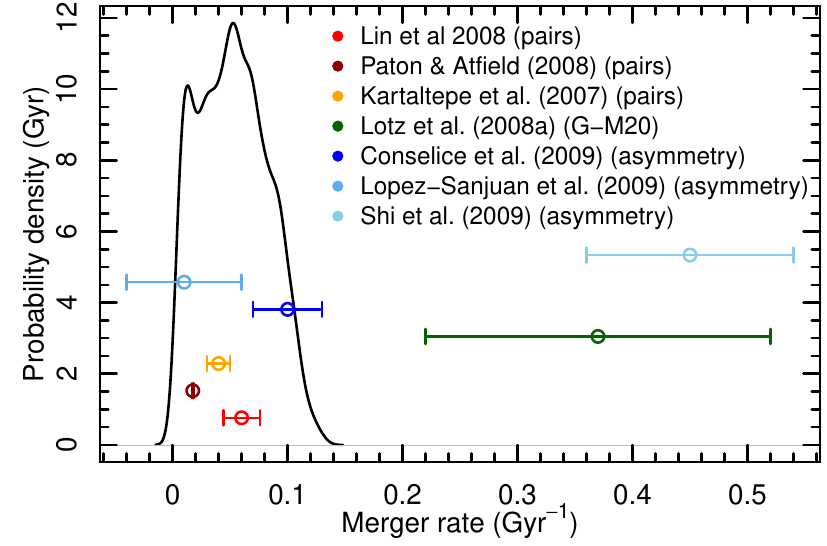}
    \caption{Posterior distribution of the major merger rate (black line), modelled for group galaxies and corrected to correspond to the whole galaxy sample (we note that Figure~\ref{fig:merge_vs_strip} shows merger rate ($\omega$) without this correction). Each point with an error bar shows a literature measurement (vertical positioning is arbitrary). Our value for $\omega = 0.11\pm0.06~{\rm Gyr^{-1}}$.}
    \label{fig:gr_merg_rate}
\end{figure}

The other stochastic event considered in our model is major merging. We found the merger rate to be $\omega = 0.11\pm0.6~{\rm Gyr^{-1}}$ which is about one major merger per $10~\rm{Gyr}$, which (compared to Hubble time) is not a major influence on the overall evolution. This result is also consistent with {conclusions by} \citet{Capozzi_2011}, who showed that the mass function of larger elliptical galaxies does not evolve considerably. The comparison of the merger rate with several literature values (\citet{Lin:2008, Patton:2008, Kartaltepe:2007, Lotz:2008, Conselice:2009, Lopez:2009, Shi:2009}) is provided in Fig.~\ref{fig:gr_merg_rate}. The present merger rate is estimated based on the data for group galaxies with the assumption that only group galaxies merge.  If we wish to compare them with all galaxies we must multiply this by the fraction of galaxies in groups, which is  about $14882/(15545+14882)\approx 0.49$, and the overall merger rate is $0.053\,{\rm Gyr^{-1}}$. We would like to point out that the relative similarity with other results is achieved via an approach that is independent of the information about galaxy pairs, morphology, and asymmetry. Our approach is independent even in calibration aspects such as pre-determining for how long the asymmetric perturbation is observable after merger.  We wish to point out that the present estimate is only an average over all clusters and galaxies, although in the case of larger sample sizes and observables we will be able to distinguish the extent to which merger rate differs in different mass clusters because merging with a galaxy that has a large relative velocity is not prone to happen \citep{Mo_book}. 
\begin{figure}
    \centering
    \includegraphics{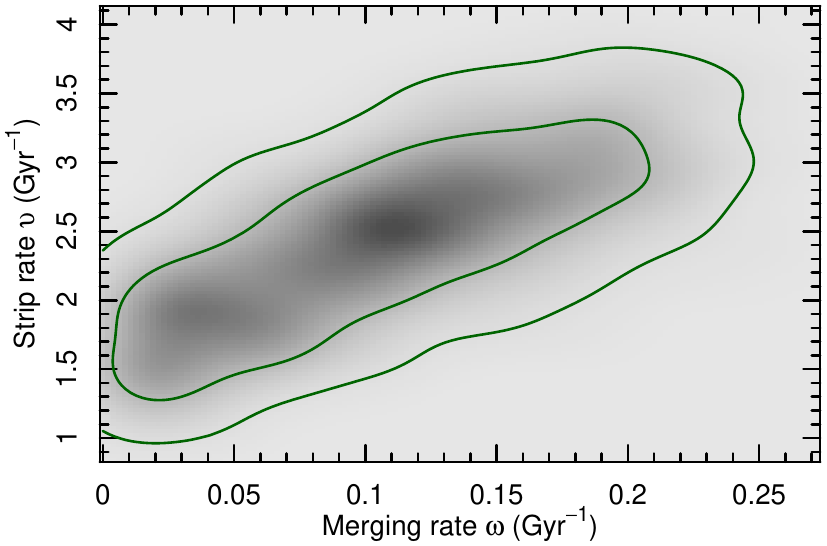}
    \caption{Correlation between merger rate and strip rate. The green lines represent $1\sigma$ and $2\sigma$ confidence intervals. The colour coding is proportional to the probability density. Point (0,0) would be no environmental effects, which is excluded. }
    \label{fig:merge_vs_strip}
\end{figure}

In our model, major mergers are accompanied with the starburst factor $s,$ which is defined as  the coefficient of SFR of merging galaxies, as described in Eq.~\eqref{Eq=:stoch_evol}. Our modelling yielded its median value, $s = 3.5,$ over the Multinest sampling. Within  one sigma interval, the value varied in the range  $s = [2.3,11]$. Major mergers are too rare for us to be able to confine $s$ more precisely from the given dataset. Considering the average efficiencies of the counter-acting processes of stripping and SFR decay, the residual SFR enhancement factor would be as low as $3.5 \times 0.8^{2.4} \times \exp(-1/1.5) = 1.05$ on average. This small value is comparable to the factor $1.2$ determined by \citet{Pearson_2019} by comparing a large sample of merged and non-merged galaxies. A more detailed comparison is difficult because in the latter work, galaxy mergers were detected long after the actual events, but in our case we consider that star formation increases instantaneously and decays rapidly. This delay is primitively mimicked by extending our calculations over the period of 1 Gyr. Another comparison can be made with \citet{dazgarca2020gas}, who estimated the SFR to be increased by a factor of $1.9\pm0.5$. As they detected mergers already in earlier stages, this value is more directly comparable to our result of $s = 3.5.$ Another example {of} an SFR increase  was estimated  by \citep{Cortijo:2017}, where they  detect an increase by a factor of between six and nine in a case study of {just} one galaxy pair. In general, we can conclude that our results on SFR enhancement in galaxy mergers are consistent with other estimates, despite the very different methodology and type of information used. 

Figure \ref{fig:merge_vs_strip} illustrates the degeneracy between ram pressure stripping and merger rate in our modelling. We see that a merging rate range as low as zero is allowed, indicating that at least in principle, late-stage galaxies can be built up without mergers, although the average value suggests that about $10\%$ of galaxies in groups merge  every gigayear. In contrast, the minimum amount of ram pressure stripping is still one event (or equivalently a $20\%$ loss of star formation) per gigayear. 

\section{Conclusions and future work}

\subsection*{Advantages and future improvements}

The main advantage of the technique used here is that it uses data as initial conditions to predict the evolution of galaxies. 
In our modelling, we combine probabilistic methods to infer galaxy properties described by analytic equations. As a result, we can include stochastic events, which is not the case for a purely analytic approach. This provides a huge advantage as we can determine the contributions from both stochastic events and smooth processes on galaxy evolution. In case of semi-analytic and numerical methods, galaxy evolution is modelled for a specific cosmology, but our modelling is independent of cosmology, except for the calibration of time from redshift. The current method is also computationally much less demanding compared to the numerical recipes adapted in hydrodynamic simulations.  We also find a high fidelity of the physical parameters that are not seen in other types of modelling. Despite the limited size of the dataset, the results we obtain is in satisfactory agreement with values from the literature, and therefore using a larger dataset or even including less massive galaxies will make for even more accurate predictions.

By using observational data for the initial conditions, the freedom of modelling is reduced,  and therefore we need fewer assumptions, and the amount of nuisance parameters is almost  non-existent. Unfortunately, this also introduces a drawback: we are unable to infer the mass or SFR distributions, but only their changes. This is because the parameters we infer are given in the  differential equations, Eq.~\eqref{eq:general_evolution}, and they describe the evolution of any galaxy property. As we are only concerned about the change in the galaxy property, the actual value does not hold relevance in our modelling. This allows us to describe the change in galaxy properties purely based on the physical processes that drive their evolution. This approach is very different from comparing galaxy properties from observations directly with  those in simulations. If a property from simulations matches very well with the observations, there is no way to break the degeneracy and point out which of the physical processes gave rise to the observed distribution. However, using this method, it is possible to distinguish the contributing process among several possibilities.

Another caveat arising due to the  use of data as initial conditions is {that we} rely heavily on the assumption that data represent the true distribution. In cases where insufficient data is available, we would fit a model of expectation to a realisation of data. 

 This paper is our first attempt at modelling galaxy evolution analytically by including both stochastic events and smooth processes. We have  covered the most relevant processes by giving equal emphasis to each process. In an upcoming paper, an  advancement will be made by including pertinent observables 
 in order to understand the relative importance of physical processes.  For example, the inclusion of observed metallicity and equations governing metallicities {will give us the} opportunity to  distinguish the amount of gas originating from pristine accretion from the gas that is being recycled. This will allow us to specify and constrain the star formation rate as there is a tight relation between metallicity, mass, and SFR \citep{LL:2010, Mannucci:2010, Lara_L_pez_2013}. {For the case where} environmental effects should be constrained, photometric observables provide a useful addition. For example, the classical bulge is thought to grow via major mergers \citep{Keselman:2012} or other processes  \citep{Bell_2017, Breda:2018}. The present framework allows us to distinguish between them.  Alternatively, if we wish to quantify the contribution of minor mergers to galaxy evolution,  we can include luminosity profiles and bars. Minor mergers influence excitation of bars \citep{ghosh2020fate}, and so we do not include them. Overall, large surveys such as J-PAS \citep{jpaspaper, minijpaspaper} or 4MOST WAVES \citep{waves:2019} {will} provide excellent datasets with which to constrain different physical processes{ to a great extent}.

\subsection*{Conclusions and summary}
In this paper we present and apply a new analytical method to isolate and quantify the contributions of different processes to galaxy quenching. The major novelty of this method is the assumption that the evolution of each galaxy  is determined by  {its} current state.  The evolution of galaxy properties is described  by a set of differential equations. These equations contain {the} evolution parameters that determine {and describe} the relative importance of the {various} processes. The knowledge of the values of the parameters comes into play when we use these differential equations to evolve or extrapolate the intermediate redshift galaxies to low redshift. As galaxies evolve along this path, we say that the parameters of the evolution are those that mimic this extrapolation most similarly along with true observations. This extrapolation is {carried out} galaxy by galaxy, but {the overall} inference {is made by} using {the} galaxy population as {a} whole. 

Using the GAMA data for galaxies with $z\lesssim 0.35$ we were able to determine the main contributions by several processes to galaxy quenching. The quenching timescale of galaxies is about $1.5\,{\rm Gyr}$ {based on modelling the evolution of isolated galaxies}. By not fixing the parameter $\beta_2$, we find that our modelling supports the idea that the  quenching speed is initially slower than the previous case and speeds up later on \citep{Maier_2019, Belli_2019}. 

We determine the {galaxy} mass {above which quenching takes place}  to be $\log_{10} M_{\rm cap}/{\rm M_\odot}\approx 10.2$. {This}  is {slightly} lower than {the} conventional {value from the literature,} $\log_{10}M/{\rm M_\odot}=10.68$. {This is because} our fit value is the starting point of mass quenching, which is natural{ly}  lower than the limit where quenching is already well observed. 

We {computed} the merger rate of galaxies, $0.053\pm0.03\,{\rm Gyr^{-1}}$, which is consistent with values from the literature. Merger rate is correlated with ram pressure stripping, and for ram pressure stripping we found that the star formation rate is reduced by $41\%$ per gigayear. 

{We have established a robust model to predict the contributions from various physical processes to galaxy quenching. The model parameters are consistent with the literature and in an upcoming paper, we will also include metallicity and galaxy photometry to make more accurate predictions. }

\section*{Acknowledgements}
 {We are grateful for the referee for the useful comments and suggestions. }This work was supported by institutional research funding  \mbox{PUTJD907} and \mbox{PRG1006} of the Estonian Ministry of Education and Research. We acknowledge the support by the Centre of Excellence "The Dark Side of the Universe" (TK133) financed by the European Union through the European Regional Development Fund. {GAMA is a joint European-Australasian project based around a spectroscopic campaign using the Anglo-Australian Telescope. The GAMA input catalogue is based on data taken from the Sloan Digital Sky Survey and the UKIRT Infrared Deep Sky Survey. Complementary imaging of the GAMA regions is being obtained by a number of independent survey programmes including GALEX MIS, VST KiDS, VISTA VIKING, WISE, Herschel-ATLAS, GMRT and ASKAP providing UV to radio coverage. GAMA is funded by the STFC (UK), the ARC (Australia), the AAO, and the participating institutions. The GAMA website is \url{http://www.gama-survey.org/} .}

\bibliography{aanda}
\appendix
\section{Characterisation of the depletion of the SFR}\label{sec:app_SF_derive}
The SFR depends on how much gas the galaxy has, or equivalently, the fraction ($f$) of gas in a galaxy that turns into stars:
\begin{equation}
    \Psi = f g^p.\label{eq:1}
\end{equation}
We denote the amount of gas in galaxy as $g$ and include the possibility that when there is too much or too little gas that there could be non-linearity $p$. 
When stars are formed, exactly the same amount of gas is removed from the galaxy, i.e.
\begin{equation}
    \Psi = -\dot{g}\label{eq:2}
.\end{equation}
By taking the first derivative of \eqref{eq:1} we get
\begin{equation}
    \dot\Psi = f p g^{p-1}\dot{g}. \label{eq:3}
\end{equation}
By solving Eq.~\eqref{eq:1} for $g$ and substituting it and Eq.~\eqref{eq:2} into Eq.\eqref{eq:3} we obtain
\begin{equation}
    \dot\Psi = -pf^{1/p}\Psi^{(2p-1)/p}.
\end{equation}
Redefining the multiplier and exponent as $\beta_1$ and $\beta_2$ we reach the form of the gas depletion in Eq.~\eqref{eq:evol_S}, given as,
\begin{equation}
    \dot\Psi = -\beta_1 \Psi^{\beta_2} .
\end{equation}

\end{document}